# Online shopping behavior study based on multi-granularity opinion mining: China vs. America

**Qingqing Zhou · Rui Xia · Chengzhi Zhang**

**Abstract** With the development of e-commerce, many products are now being sold worldwide, and manufacturers are eager to obtain a better understanding of customer behavior in various regions. To achieve this goal, most previous efforts have focused mainly on questionnaires, which are time-consuming and costly. The tremendous volume of product reviews on e-commerce websites has seen a new trend emerge, whereby manufacturers attempt to understand user preferences by analyzing online reviews. Following this trend, this paper addresses the problem of studying customer behavior by exploiting recently developed opinion mining techniques. This work is novel for three reasons. First, questionnaire-based investigation is automatically enabled by employing algorithms for template-based question generation and opinion mining-based

Qingqing Zhou

Department of Information Management, Nanjing University of Science and Technology, Nanjing, 210094, China

Alibaba Research Center for Complex Sciences, Hangzhou Normal University, Hangzhou, 311121, China

Rui Xia

School of Computer Science and Engineering, Nanjing University of Science and Technology, Nanjing, 210094, China

Chengzhi Zhang (corresponding author)

Department of Information Management, Nanjing University of Science and Technology, Nanjing, 210094, China

Alibaba Research Center for Complex Sciences, Hangzhou Normal University, Hangzhou, 311121, China

Jiangsu Key Laboratory of Data Engineering and Knowledge Service, Nanjing University, Nanjing, 210093, China

Tel: +86-025-84315963

E-mail: zhangcz@njust.edu.cn

| ★★☆☆☆ **Not Happy**<br>Battery life is very disappointing.<br>Uses at least 50% charge a day<br>without a lot of use! | ★★★★★正品，质量好，快递迅速<br>外包装规矩，正品，质量有保证。亚马逊快递迅速，服务好<br>(The package is proper, good product, guaranteed quality, Fast express mail by Amazon, good service) |
|---|---|
| (a) English review | (b) Chinese review |

**Fig. 1** Examples of English and Chinese reviews on Apple iPhone 5s.

answer extraction. Using this system, manufacturers are able to obtain reports of customer behavior featuring a much larger sample size, more direct information, a higher degree of automation, and a lower cost. Second, international customer behavior study is made easier by integrating tools for multilingual opinion mining. Third, this is the first time an automatic questionnaire investigation has been conducted to compare customer behavior in China and America, where product reviews are written and read in Chinese and English, respectively. Our study on digital cameras, smartphones, and tablet computers yields three findings. First, Chinese customers follow the Doctrine of the Mean, and often use euphemistic expressions, while American customers express their opinions more directly. Second, Chinese customers care more about general feelings, while American customers pay more attention to product details. Third, Chinese customers focus on external features, while American customers care more about the internal features of products.

# 1 Introduction

With the rapid development of e-commerce, more and more businesses are operating on a global scale. For manufacturers, customer behavior plays a key role in decision-making regarding product design and marketing. It is also important for transactional enterprises to conduct questionnaire-based surveys of customer behavior. Most existing research on customer behavior is conducted via manually administered questionnaires, which is expensive, time-consuming, and prone to error. Nowadays, there is a tremendous volume of product reviews available on e-commerce websites, and manufacturers are attempting to understand user preferences by accessing online customer reviews. For example, C&A Marketing [1] analyzes product reviews on Amazon to understand users' requirements, and this information is then used to design new products. To this end, product reviews can reveal the attitudes and preferences of customers in different regions. Two illustrative reviews of the *Apple iPhone 5s* are shown in Figure 1. As can be seen, the customer in America is unhappy with the battery, while the customer in China is satisfied with the packaging and service. These reviews are typical among American and Chinese consumers, respectively. In general, there are numerous differences between customers in different regions, who usually focus on different aspects of products and display different attitudes in relation to the same aspects. To identify these differences, we exploit recently developed opinion mining techniques to complete questionnaires automatically based on reviews written by Chinese and American customers.

---
[1] http://www.camarketing.com

We generate questions automatically based on templates, and extract answers by sentiment analysis. 1) **For question generation**, we build question templates for overall sentiment, brand preference, and purchase decision factors, respectively. 2) **For answer extraction**, we use the topic model for aspect extraction, and we conduct multi-granularity sentiment analysis to identify sentiment polarities.

Using three IT product domains as a case study, we aim to identify preferences of customers in different regions. Three interesting results are obtained from the perspective of sociology, based on the empirical results:

(1) Chinese customers obey the Doctrine of the Mean, while American customers express opinions more directly

(2) Chinese customers pay more attention to general feelings, while American customers focus on product details

(3) Chinese customers are more concerned with external features, while American customers pay more attention to internal features.

The remainder of this paper is organized as follows. In Section 2, we review related work. Data collection is introduced in 3. In Section 4, we present our methodology for automatic question generation and answer extraction. Section 5 presents our comparisons and analysis. Section 6 concludes.

## 2 Related work

There are three areas of research related to our study: online shopping behavior, automatic question and answer generation, and sentiment analysis.

### 2.1 Online shopping behavior study

In the field of e-commerce, many studies have focused on online shopping behavior. Chatterjee examined the effect of negative reviews on retailer evaluation and patronage intention, given that the consumer has already made a product/brand decision [9].

The results indicated that the extent of their word-of-mouth search depended on the consumers' reasons for choosing an online retailer. Park et al. proved that the quality of online reviews had a positive effect on customers' purchase intention, and that purchase intention increased as the number of reviews increased [41]. They also found that low-involvement consumers were affected by the quantity of reviews rather than review quality, but high-involvement consumers were mainly affected by review quantity when the review quality was high. Lee et al. investigated the effects of negative online consumer reviews on consumers' attitudes to products [21]. The experimental results showed that a high proportion of negative online consumer reviews elicited a conformity effect.

Vermeulen and Seegers applied consideration set theory to model the impact of online hotel reviews on consumer choice [49]. They proposed an experimental study that included review valence, hotel familiarity, and reviewer expertise as independent factors. The results showed that on average, exposure to online reviews enhanced hotel consideration among consumers. Park and Kim showed that the effect of cognitive fit (types of reviews) on purchase intention was stronger for experts than for novices, while the effect of the number of reviews on purchase intention was stronger for novices than for experts [40]. Zhu and Zhan examined how product and consumer characteristics moderated the influence of online consumer reviews on product sales using data from the video game industry [62]. Their findings indicated that online reviews were more influential for less popular games and games whose players had more Internet experience.

As a result of the emergence of e-commerce and the development of global markets, researchers have begun to compare customers' online shopping preferences and behaviors in different cultures [35]. Mahmood et al. conducted an online shopping behavior study using data from 26 nations and analyzed constructs using a structural equation model [31]. They found that trust and economic conditions, but not educational level and technological savvy, had a significant positive effect on online shopping behavior. Hwang et al. investigated online shopping preferences in three countries, the US, Korea, and Turkey [18]. They found significant differences in online shopping preferences, especially in relation to information accuracy, security, and price comparison.

From the above, it can be seen that existing studies of customer behavior can be divided into two categories. In the first category, studies have focused on coarse-grained review mining, i.e., only sentiment polarities of reviews are used to conduct studies of customer behavior. In the second category, comparative studies are conducted on customers' online shopping behavior. These are mainly based on traditional survey methods such as questionnaires. These methods have several disadvantages, including small sample sizes, long cycles, and high costs. In summary, existing studies are mainly focused on relatively macroscopic online shopping behavior rather than focusing on specific products or differences in customer attitudes. In contrast, our study compares online shopping behavior by generating questions and answers automatically via multi-granularity review mining.

## 2.2 Automatic question and answer generation

In response to the abovementioned disadvantages of traditional methods, research on automatic question and answer generation has attracted the interest of many researchers. Sumita et al. proposed automatic generation of fill-in-the-blank questions (FBQs) together with testing based on item response theory (IRT) to measure English proficiency [45]. Their method involved three steps. First, an FBQ was generated from a given sentence in English. Second, each of the candidates for incorrect choices was verified using the Web. Finally, the blank sentence, the correct choice, and the incorrect choices surviving verification were arranged to form the FBQ. Brown et al. described an approach to automatically generate questions for vocabulary assessment [6]. Unlike traditional handwritten assessments, they generated six types of vocabulary questions, including word-bank and multiple-choice questions. Their experimental results suggested that these automatically generated questions gave a measure of vocabulary skill that correlated well with subject performance on independently developed manually prepared questions.

The lack of automatic methods for evaluating answers is an impediment to progress in the field of question and answer generation. Lin et al. proposed a measure called Pourpre for automatically evaluating answers to complex questions based on n-gram co-occurrences between machine output and a human-generated answer key [27]. Based on the engagement taxonomy and benefits of question answering during the algorithm visualization, Myller and Niko proposed incorporating automatic question generation into a program visualization tool, Jeliot 3 [36]. avidov et al. presented a novel framework for the discovery and representation of general semantic relationships that exist between lexical items [13]. To assess the quality of relationships, they used pattern clusters to automatically generate Scholastic Assessment Test analogy questions. Papasalouros et al. presented an innovative approach to generating multiple-choice questions automatically [39]. Their method was based on domain-specific ontologies and was independent of lexicons or other linguistic resources. With the increasing popularity of community-based question-answering services such as Yahoo! Answers, huge numbers of user-generated questions are now available. Lin used this data along with search engine query logs to create a shared question generation task that aimed to automatically generate questions given a query [26]. Liu et al. presented a novel automatic question generation approach that generated trigger questions as a form of support for students learning through writing [29]. Experimental results showed that the questions generated by their system significantly outscored generic questions on overall quality measures.

From the above, we can see that existing studies are either about information retrieval or about generating tests for skills upgrading. There are few studies focusing on questionnaire generation from online reviews. Our work generates questions and extracts answers automatically to compare online behavior via sentiment analysis of online reviews.

## 2.3 Sentiment analysis

To identify answers automatically, we conducted multi-granularity sentiment analysis at both the document level and the aspect level. Document-level sentiment analysis aims to determine whether a document expresses a sentiment (i.e., positive or negative) [59, 28]. Aspect-level sentiment analysis performs finer-grained analysis to identify various aspects and the sentiment polarities that people express in relation to them [19, 60].

### 2.3.1 Document-level sentiment analysis

Document-level sentiment analysis is used to predict whether the document as a whole expresses a positive sentiment or a negative one [59]. This can be achieved by using one of two learning methods: supervised and unsupervised. In our case, many existing supervised learning methods can be applied [38]. For example, Mullen and Collier used Support Vector Machine (SVM) to conduct sentiment analysis, which brought together diverse sources of potentially pertinent information [34]. To improve the accuracy of sentiment classification, some special techniques for sentiment analysis have been proposed, including those by Xia et al. [54, 56, 53, 55], Li et al. [23], Nakagawa et al. [37] and Maas et al. [30]. One of the most classic unsupervised learning approaches to sentiment analysis was proposed by Turney et al. [48]. In their work, they compiled syntactic patterns that people used to express opinions based on parts of speech, and then performed sentiment analysis measuring the pointwise mutual information from these patterns, which is effcient. Other unsupervised learning methods such as those of Taboada et al. [46] and Denecke & Kerstin [14] were based on sentiment lexicons. Chikersal et al. described a Twitter sentiment analysis system that classified a tweet as either positive or negative based on its overall tweet-level polarity [10]. The proposed method was evaluated using two publicly available Twitter corpora to illustrate its effectiveness.

### 2.3.2 Aspect-level sentiment analysis

Rather than providing isolated opinions about a product, customers tend to compare specific features of products. Therefore, it is necessary to conduct aspect-level sentiment analysis [28], including aspect extraction and aspect sentiment identification [60, 15]. Many studies of aspect extraction have been carried out. Popescu and Etzioni improved the mining methods via pointwise mutual information to extract product features and opinions from reviews [42]. Jo and Oh proposed the Sentiment-LDA model and the aspect and sentiment unification model (ASUM) for reviews of electronic devices and restaurants [19]. Their results showed that ASUM outperformed other generative models and performed similarly to supervised classification models. Moghaddam and Este developed some design guidelines for aspect-based opinion mining by discussing a series of increasingly sophisticated LDA models [33].

For aspect sentiment identification, Ding et al. proposed a simple method based on a lexicon, which was shown to be quite effective in a large number of domains [15]. Blair-Goldensohn et al. [3] and Yu et al. [60] used similar methods to identify aspect sentiments. Wei and Gulla proposed a hierarchical classification model to determine the scope of each sentiment expression [51]. They also presented an approach for generating multi-unigram features to enhance a negation-aware naïve Bayes classifier for sentiment analysis of sentences in product reviews [52]. Thet et al. performed fine-grained analysis to determine both the orientation and strength of a reviewer's sentiments in relation to various aspects of a movie [47]. Gangemi et al. inferred the conceptual and affective information associated with opinions [16]. To solve contextual polarity ambiguity, Cambria et al. focused on the use of semantics to better explore opinion-level context [7]. Xia et al. explored opinion-level context to solve contextual polarity ambiguity [57]. Das and Bandyopadhyay proposed tagging sentence-level emotions and designed a valence based on the word-level constituents on the SemEval 2007 affect sensing news corpus [12]. Cambria et al. introduced novel statistical approaches to concept-level sentiment analysis that went beyond a mere syntax-driven analysis of text and provided a semantics-based method [8]. Li et al. implemented a generic stock-price prediction framework by analyzing the impact of news reports on stock prices via sentiment analysis [24]. Rill et al. presented a system called PoliTwi, which was designed to detect emerging political topics earlier on Twitter than on other standard information channels [43]. Agarwal et al. focused on finding good feature sets to conduct concept-level sentiment analysis, and achieved excellent performance [1].

**Table 1** Statistics of our review corpora.

| Products | Brands | #Chinese reviews | #English reviews |
|----------|--------|------------------|------------------|
| Digital camera | Cannon | 2,245 | 2,550 |
| | Nikon | 2,172 | 2,419 |
| Smart phone | Apple | 1,805 | 3,578 |
| | Samsung | 2,579 | 5,525 |
| Tablet computer | Apple | 485 | 3,189 |
| | Samsung | 253 | 927 |

In this study, we conduct multi-granularity sentiment analysis to generate questions and answers automatically to identify review-based differences between customers from different regions. Specifically, we use a supervised machine learning method to conduct document-level sentiment analysis and a lexical affinity method for aspect-level sentiment analysis. We aim to identify differences between opinions expressed by customers in different regions from the perspective of sociology. To the best of our knowledge, this is the first work to compare the online shopping behavior of Chinese and American customers from this perspective via automatic question and answer generation.

## 3 Data collection

This study aims to compare the behavior of online customers in China and America. Therefore, we decided to collect reviews on the same products from Amazon.com and Amazon.cn, respectively. Note that the reviews from Amazon.cn are mostly written in Chinese, while those from Amazon.com are mostly written in English. The sample includes reviews of three kinds of IT products: digital cameras (Canon and Nikon), smartphones (Apple iPhone and Samsung Galaxy) and tablet computers (Apple iPad and Samsung Galaxy) [2]. We collected 10,498 Chinese reviews and 18,498 English reviews from 7th October, 2009 to 5th August, 2014. After deleting low-quality reviews, including reviews with no text, garbled reviews, and advertisements, 9539 Chinese reviews and 18,188 English reviews were included in our sample. Statistics on the sample of reviews are presented in Table 1.

To identify customers' attitudes towards the products, we used the supervised machine learning method to conduct sentiment analysis. Therefore, we constructed a training sample by manually tagging sentiment polarities of 1800 randomly selected Chinese reviews and 3600 randomly selected English reviews. Statistics on these manually annotated reviews are presented in Table 2.

---

[2] We select the brands with the most two quantity of reviews from Amazon as our data set.

**Table 2** Statistics on the annotated review corpora.

|  | #Chinese reviews | #English reviews |
|---|---|---|
| **#positive** | 1,200 | 2,400 |
| **#negative** | 600 | 1,200 |
| **sum** | 1,800 | 3,600 |

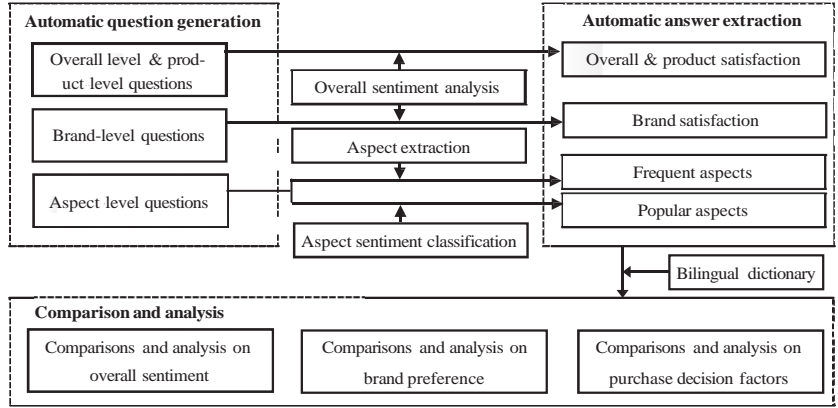

**Fig. 2** Framework of opinion comparison between Chinese and American customers.

## 4 Methodology

### 4.1 Framework

In this study, our aim is to identify the cognitive differences between Chinese and American customers by automatic question and answer extraction, which is based on opinion mining. The framework, which is shown in Figure 2, includes three major parts.

1) **Automatic question generation**. We generate questions automatically based on templates, which includes overall & product-level, brand-level and aspect-level questions.

2) **Automatic answer extraction**. We extract answers for the three levels using multi-granularity sentiment analysis. Specifically, we use overall sentiment analysis to extract answers about overall & product satisfaction and brand satisfaction, and we use aspect-level sentiment analysis to extract answers about frequent aspects and popular aspects.

3) **Comparison and analysis**. We use a bilingual dictionary [3] to map Chinese and English aspects prior to comparisons. We analyze differences between Chinese and

---

[3] https://catalog.ldc.upenn.edu/LDC2002L27

**Table 3** Question template for overall sentiment (using digital cameras as an example).

| Do you like _____ ? | | |
|---|---|---|
| **Level** | **Question** | |
| **Overall satisfactions** | Do you like *digital products* | ? |
| **Product satisfactions** | Do you like *digital cameras* | ? |

**Table 4** Question template for brand preference (using digital cameras as an example).

| Which brand of _____ do you prefer? | | | | |
|---|---|---|---|---|
| Which brand of *digital cameras* | do you prefer? | A. *Cannon* | B. *Nikon* | |

American customers at three levels: overall sentiment, brand preferences, and purchase decision factors.

The methods used for automatic question generation and answer extraction are described in detail in Section 4.2 and 4.3, respectively.

## 4.2 Automatic question generation

We automatically generate questions using manually constructed templates. There are three subtasks in this part: 1) overall sentiment, 2) brand preference, and 3) purchase decision factors, which are elaborated as follows.

(1) Generating questions of overall sentiment

One of our goals was to compare the differences between Chinese and American customers in relation to overall & product satisfaction. Therefore, we designed the question template shown in Table 3. This template includes overall-level and product-level questions. Regarding overall satisfaction, the blank space is filled with '*digital products*'. For product-level satisfaction, the blank space is filled with either '*digital cameras*', '*smartphones*', or '*tablet computers*'.

(2) Generating questions of brand preference

To compare the differences in brand preference, the question template is '*Which brand of do you prefer*?' as shown in Table 4. The blank space is filled with either '*digital camera*', '*smartphones*', or '*tablet computers*'. Each question has two options.

(3) Generating questions of purchase decision factors

To extract frequent and popular aspects, respectively, we designed two question templates: 1) '*What aspects of do you care about*?' and 2) '*What aspects of ____ are you satisfied with*?' as shown in Table 5. The blank spaces are filled with either '*digital cameras*', '*smartphones*', or '*tablet computers*'.

**Table 5** Question templates for purchase decision factors (using digital cameras as an example).

| **Frequent** | **aspect** | What aspects of | *digital cameras* | do you care about? |
|---|---|---|---|---|
| **Popular** | **aspect** | What aspects of | *digital cameras* | are you satisfied with? |

### 4.3 Automatic answer extraction

In previous studies, comparisons of online shopping behavior were mainly conducted using traditional questionnaires, which is a costly and time-consuming process. In this study, we conduct the comparison by generating questionnaires and extracting answers automatically. Specifically, we extract answers using opinion mining, including overall sentiment classification, aspect extraction, and aspect sentiment classification. There are three subtasks in this part: 1) extracting answers regarding overall sentiment, 2) extracting answers regarding brand preference, and 3) extracting answers regarding purchase decision factors.

### (1) Extracting answers of overall sentiment

Sentiment classification is a hot topic in natural language processing. It can be used to analyze online public opinion, filter noise. Document-level sentiment analysis is used to predict whether the review as a whole expresses a positive sentiment or a negative one. To extract answers regarding overall and product satisfaction, we conduct overall sentiment classification [4]. Specifically, we used linear SVM [11] as the classification model with CHI [20] as the feature selection method, and Term Frequency-Inverse Document Frequency (TF-IDF) [44] as the feature weighting method. In addition, we use Jieba [5] for Chinese word segments and LibSVM as the classifier. For overall satisfaction, we first conduct overall sentiment classification to identify sentiment polarity in reviews. Then, we count the numbers of positive and negative reviews. Finally, we obtain a value for overall satisfaction via the value for overall sentiment (OS) , which can be calculated via equation 1:

$$OS = \frac{\#pos}{\#pos + \#neg}$$

(1)

where *#pos* denotes the number of positive reviews, and *#neg* denotes the number of negative reviews. The same method is also used in relation to product satisfaction.

For example, to answer the question '*Do you like digital products*?' we conduct overall sentiment analysis to obtain OS values. The OS values for the Chinese samples and English samples are 0.8873 and 0.8436, respectively. Therefore, it can be seen that most customers are satisfied with digital products.

---

[4] Overall sentiment classification here is similar to document-level sentiment classification

[5] https://github.com/fxsjy/jieba

(2) Extracting answers of brand preference

To obtain corresponding results in relation to brand satisfaction, first, we conduct overall sentiment classification to identify the sentiment polarities of reviews about brands A and B. Then, we count the number of reviews for each brand. Finally, we obtain brand preference (BP) values using equation 2:

$$BP = \begin{cases} A, & \text{if } OS(A) > OS(B) \\ B, & \text{otherwise} \end{cases} \quad (2)$$

where $OS(A)$ and $OS(B)$ can be calculated via equation 3:

$$OS(B) = \frac{\#pos(X)}{\#pos(X) + \#neg(X)} \quad (3)$$

where $OS(X)$ denotes OS value of brand X, $\#pos(X)$ denotes the number of positive reviews about brand X, and $\#neg(X)$ means the number of negative reviews about brand X.

For example, the answer to the question '*Which brand of smartphone do you prefer?*' has two options, *Apple* and *Samsung*. We find that $OS(Apple)$ is 0.6401 and $OS(Samsung)$ is 0.7548, hence the BP is Samsung.

(3) Extracting answers of purchase decision factors

To find frequent and popular aspects, we use two processes: 1) aspect extraction and 2) aspect sentiment classification.

Aspect extraction is used to extract frequent aspects. The topic model method is widely used for aspect extraction [25, 61]. In addition, when people comment on different aspects of an entity, the vocabulary that they use generally converges. Thus, nouns that appear frequently are usually genuine and important aspects [38]. In this study, we combine the two methods to extract aspects. Specifically, we first tag the parts of speech for words in each topic generated by LDA using Jieba [6] for Chinese words and NLTK [7] for English words, and then choose nouns as candidate aspects extracted with LDA [4]. We calculate the sum of probabilities for each noun in every topic, and choose the highest ones as candidate aspects. Then, we use frequent aspect (FA) values to find more candidate aspects, which can be calculated using equation 4:

$$FA(Y) = \#pos(Y) + \#neg(Y) \quad (4)$$

---

[6] https://github.com/fxsjy/jieba

[7] http://www.nltk.org/

where #*pos*(*Y*) denotes the numbers of positive reviews about aspect Y, and #*neg*(*Y*) means the numbers of negative reviews about aspect Y. Finally, we extract the most frequent aspects by sorting the FA values of all candidate aspects.

For example, for the question '*What aspects of digital cameras do you care about?*', we first use the combined method to extract frequent aspects about *digital cameras*. Then, we calculate the FA value for each frequent aspect. FA(package) is 709 and FA(genuine product) is 684 in the Chinese samples, which are much higher than the FA values of other aspects. Therefore, it can be seen that Chinese customers care more about *package* and whether it is a *genuine product* than about other aspects.

Aspect sentiment classification is used to detect attitudes toward a given aspect. For each review, we first identify sentiment words via sentiment lexicons in Chinese [8]and English [9]. It is difficult to determine which aspect a sentiment word is describing when a review expresses sentiments on multiple aspects. In general, if the distance between an aspect word and a sentiment word is short, the sentiment word is more likely to describe the aspect. Therefore, we compute the sentiment polarity of each aspect in a review by measuring the distance between an aspect word and a sentiment word [15]. Sentiment polarity (SP) of aspect *Y* in review *s* can be calculated using equation 5:

$$SP(Y, s) = \begin{cases} +1, & if \sum_{w_i : w_i \in s \cap w_i \in V} \frac{w_i.SO}{dis(w_i, Y)} \\ -1, & otherwise \end{cases} \tag{5}$$

where $w_i.SO$ is calculated by equation 6:

$$w_i.SO = \begin{cases} +1, & if \ w_i \ is \ a \ positive \ word, \\ -1, & otherwise \end{cases} \tag{6}$$

where $w_i$ denotes a sentiment word, *V* means the set of all sentiment words, *s* is the review that contains the aspect *Y*, and $dis(w_i, Y)$ denotes the distance between aspect Y and sentiment word $w_i$ in review *s*. The sentiment score of the word $w_i$ is denoted by $w_i.SO$.

To find popular aspects, we first conduct aspect sentiment classification to calculate SP values of aspects in each review. Then, we use popular aspect (PA) values to extract popular aspects, which can be calculated using equation 7:

$$PA(Y) = \frac{\sum_{i=1}^{m}(SP(Y, s_i) + 1)}{2 \sum_{i=1}^{m} |SP(Y, s_i)|} \tag{7}$$

where *m* denotes the number of reviews about aspect *Y*.

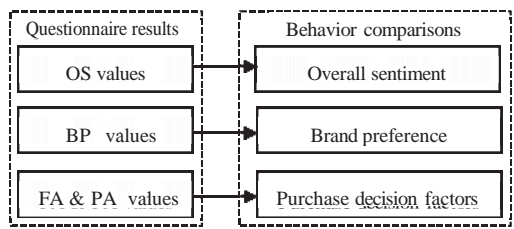

**Fig. 3** The flowchart of empirical study for online shopping behavior.

For example, for the question '*What aspects of digital cameras are you satisfied with*?', PA(quality) is 0.9900 and PA(flashlight) is 0.9611 in English samples, which are much higher than PA values of other aspects. Therefore, it can be seen that English customers are more satisfied with the overall quality and the flashlight than with other aspects.

## 5 Empirical study

We conducted our analysis using digital cameras, smartphones, and tablet computers. The process is shown in Figure 3. The results of our investigation are presented in Section 5.1. Based on the analysis in Section 5.1, we present the results of our analysis of online consumer behavior in Section 5.2. Finally, in order to illustrate the value of our proposed method, we present an example of a smartphone promotion in Section 5.3.

### 5.1 Questionnaire investigation results

We obtained our results using multi-granularity sentiment analysis. There are three types of results: 1) overall sentiment, 2) brand preference, and 3) purchase decision factors, which are elaborated as follows.

#### 5.1.1 Overall sentiment

The results of the analysis of overall sentiment include overall satisfaction and product satisfaction. Regarding overall satisfaction, the question we asked was '*Do you like digital products*?'. Our sample consisted of 9539 Chinese reviews and 18,188 English reviews. Table 6 shows the results in terms of overall satisfaction. It can be seen that the OS value in Chinese reviews is higher than that in English reviews, which indicates that Chinese customers' satisfaction regarding digital products is higher than that of American customers.

**Table 6** Results of overall sentiment.

|                    | #Review | #Pos.  | #Neg. | OS     |
|--------------------|---------|--------|-------|--------|
| **Chinese reviews** | 9,539   | 8,341  | 1,198 | 0.8873 |
| **English reviews** | 18,188  | 14,763 | 3,425 | 0.8436 |

**Table 7** Results of product sentiment.

|                    | Products        | #Review | #Pos.  | #Neg. | OS     |
|--------------------|-----------------|---------|--------|-------|--------|
| **Chinese reviews** | Camera          | 4,417   | 4,054  | 363   | 0.9178 |
|                    | Smartphone      | 4,384   | 3,607  | 777   | 0.8227 |
|                    | Tablet computer | 738     | 680    | 58    | 0.9214 |
| **English reviews** | Camera          | 4,969   | 4,697  | 272   | 0.9452 |
|                    | Smartphone      | 9,103   | 6,460  | 2,643 | 0.7096 |
|                    | Tablet computer | 4,116   | 3,606  | 510   | 0.8761 |

For product satisfaction, the question we asked was '*Do you like digital camera / smartphone / tablet computer*?'. We collected 4417 Chinese reviews and 4969 English reviews about digital cameras, 4384 Chinese reviews and 9103 English reviews about smartphones, and 738 Chinese reviews and 4116 English reviews about tablet computers. Table 7 shows the results in relation to product satisfaction. For digital cameras, the Chinese OS value is 91.78%, while the English OS value is 94.52%. For smartphones, the Chinese OS value is 82.27%, while the English OS value is 70.96%. For tablet computers, the Chinese OS value is 92.14%, while the English OS value is 87.61%. The results indicate that 1) for digital cameras, both Chinese and American customers are satisfied with the cameras, although the American customers' satisfaction tends to be slightly higher, and 2) for smartphones and tablet computers, the OS values of American customers are significantly lower than those of Chinese customers.

### 5.1.2 Brand preference

The results in relation to brand preference are shown in Figure 4. OSC denotes the Chinese OS value for each brand, while OSE denotes the English OS value for each brand. For digital cameras, we collected 4795 reviews of Canon products (2245 Chinese reviews and 2550 English reviews) and 4591 reviews of Nikon products (2172 Chinese reviews and 2419 English reviews). For smartphones, we collected 5383 reviews of Apple products (1805 Chinese reviews and 3578 English reviews) and 8104 reviews of Samsung products (2579 Chinese reviews and 5525 English reviews). For tablet computers, we collected 3674 reviews of Apple products (485 Chinese reviews and 3189 English reviews) and 1180 reviews of Samsung products (253 Chinese reviews and 927 English reviews). It can be seen from Figure 4 that the differences between the OSCs are smaller than those between the OSEs in all three domains, especially in the smartphone domain, which indicates that American customers have more obvious brand preferences than Chinese customers.

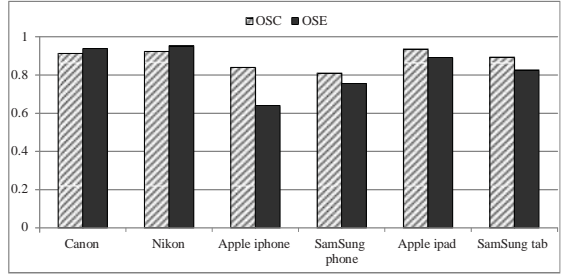

**Fig. 4** OS values of different brands.

### 5.1.3 Purchase decision factors

To extract purchase decision factors, we performed three subtasks: 1) extracting the most frequent aspects, 2) extracting the most popular aspects, and 3) extracting frequent aspects that are a concern of both Chinese and American customers.

(1) Most frequent aspect extraction and comparison

In this study, we use gensim [10] to identify topics with both aspects and sentiments. The LDA algorithm in gensim is based on Hoffman et al. [17]. They developed an online variational Bayes algorithm for LDA. We use the POS taggers Jieba and NLTK to perform Chinese and English POS tagging, respectively. We choose nouns extracted by LDA as candidate aspects, and then choose the top 10 aspects in terms of aspect frequency as real aspects via FA values. For the question '*Which aspects of digital camera / smartphone / tablet computer do you care about / are you satisfied with*?', we collected 3280 Chinese reviews and 7098 English reviews about digital cameras, 4040 Chinese reviews and 21,798 English reviews about smartphones, and 1474 Chinese reviews and 8780 English reviews about tablet computers. The 10 most frequent aspects of digital cameras, smartphones, and tablet computers are shown in Table 8(a), Table 9(a) and Table 10(a) respectively. We use the entropy value to measure the differences between Chinese and American customers regarding the 10 most frequent aspects, which can be calculated using equation 8:

$$E_{fre} = -\sum_{i=1}^{N_{fre}}\left(\frac{FA_i}{\sum_{i=1}^{N_{fre}}FA_i} * \log \frac{FA_i}{\sum_{i=1}^{N_{fre}}FA_i}\right) \qquad (8)$$

---

[10] http://radimrehurek.com/gensim/

**Table 8** Top aspects of digital cameras.

| (a)Top frequent aspects | | | | (b)Top popular aspects | | | |
|---|---|---|---|---|---|---|---|
| **Chinese Reviews** | | **English Reviews** | | **Chinese Reviews** | | **English Reviews** | |
| Aspects | FA | Aspects | FA | Aspects | PA | Aspects | PA |
| package | 709 | lens | 2,118 | genuine product | 0.9985 | quality | 0.9900 |
| genuine product | 684 | quality | 1,604 | logistics | 0.9295 | flashlight | 0.9611 |
| price | 488 | price | 994 | price | 0.9262 | battery | 0.9402 |
| lens | 467 | lighter | 950 | package | 0.9027 | price | 0.9054 |
| logistics | 312 | screen | 406 | quality | 0.8901 | lens | 0.8820 |
| quality | 273 | sensor | 272 | appearance | 0.8772 | appearance | 0.8720 |
| battery | 149 | button | 270 | lens | 0.8287 | screen | 0.8670 |
| service | 82 | appearance | 250 | battery | 0.7987 | button | 0.8593 |
| screen | 59 | battery | 234 | screen | 0.7288 | sensor | 0.8382 |
| appearance | 57 | shutter | 188 | service | 0.6829 | shutter | 0.8298 |

Where $FA_i$ means frequent aspect (FA) value of the $i$th aspect, which can be calculated by equation 4, $N_{fre}$ denotes the number of frequent aspects. Table 11 shows the entropy values of Chinese and American customers regarding frequent aspects. From Table 11, it can be seen that Chinese customers' concerns about frequent aspects are less homogeneous than those of American customers in two of the three domains. Hence, we can conclude that concerns about frequent aspects are obviously different in all three domains between Chinese and American customers. Specifically, from Table 8(a), we can see that for digital camera reviews, the most frequent aspect in Chinese reviews is *package*, followed by *genuine product* and *price*, while American customers are most concerned about the *lens*, followed by *quality* and *price*. From Table 9(a), we can see that for smartphone reviews, the most frequent aspect in Chinese reviews is *logistics*, followed by *genuine product* and *screen*, while American customers are most concerned about the *apps*, followed by *screen* and *battery*. From Table 10(a), we can see that for tablet computer reviews, the most frequent aspect in Chinese reviews is *screen*, followed by *logistics* and *price*, while American customers are most concerned about *apps*, followed by *screen* and *appearance*.

(2) Most popular aspect extraction and comparison

The 10 most popular aspects for digital cameras, smartphones, and tablet computers are shown in Table 8(b), Table 9(b) and Table 10(b) respectively. We also use the entropy value to measure the differences between Chinese and American reviews regarding the 10 most popular aspects, which can be calculated using equation 9:

$$E_{pop} = -\sum_{i=1}^{N_{pop}}(PA_i * logPA_i) \tag{9}$$

**Table 9** Top aspects of smartphones.

| (a)Top frequent aspects | | | | (b)Top popular aspects | | | |
|---|---|---|---|---|---|---|---|
| **Chinese Reviews** | | **English Reviews** | | **Chinese Reviews** | | **English Reviews** | |
| Aspects | FA | Aspects | FA | Aspects | PA | Aspects | PA |
| logistics | 760 | app | 6910 | genuine product | 0.9848 | quality | 0.97198 |
| genuine product | 726 | screen | 3200 | price | 0.8697 | apperance | 0.9154 |
| screen | 494 | battery | 2458 | logistics | 0.8592 | camera | 0.8747 |
| service | 481 | apperance | 2034 | quality | 0.813 | battery | 0.8373 |
| appearance | 391 | price | 1538 | appearance | 0.7928 | charger | 0.8363 |
| price | 376 | camera | 1516 | screen | 0.7874 | screen | 0.8363 |
| battery | 274 | service | 1164 | service | 0.7775 | price | 0.8349 |
| quality | 246 | quality | 1142 | app | 0.6918 | service | 0.8196 |
| app | 159 | charger | 1014 | memory | 0.6917 | button | 0.7981 |
| memory | 133 | button | 822 | battery | 0.6496 | app | 0.737 |

**Table 10** Top aspects of tablet computers.

| (a)Top frequent aspects | | | | (b)Top popular aspects | | | |
|---|---|---|---|---|---|---|---|
| **Chinese Reviews** | | **English Reviews** | | **Chinese Reviews** | | **English Reviews** | |
| Aspects | FA | Aspects | FA | Aspects | PA | Aspects | PA |
| screen | 512 | app | 2385 | service | 0.95556 | quality | 0.9969 |
| logistics | 305 | screen | 1584 | genuine product | 0.8385 | size | 0.9654 |
| price | 145 | size | 1473 | battery | 0.8356 | service | 0.9466 |
| genuine product | 130 | price | 832 | appearance | 0.8305 | resolution | 0.945 |
| resolution | 85 | battery | 648 | button | 0.7714 | price | 0.9447 |
| quality | 85 | quality | 636 | resolution | 0.7647 | battery | 0.9414 |
| battery | 73 | resolution | 400 | quality | 0.7177 | button | 0.9317 |
| appearance | 59 | processor | 282 | price | 0.7103 | app | 0.8973 |
| service | 45 | button | 278 | logistics | 0.7049 | screen | 0.8813 |
| button | 35 | service | 262 | screen | 0.6191 | processor | 0.8511 |

**Table 11** Entropy values about frequent aspects.

| | **Digital cameras** | **Smart phones** | **Tablet computers** |
|---|---|---|---|
| Chinese | 0.8795 | 0.9436 | 0.8415 |
| English | 0.8495 | 0.8959 | 0.8832 |

Where $PA_i$ means PA value of the $i$th aspect, which can be calculated using equation 7, and $N_{pop}$ denotes the number of popular aspects. Table 12 shows the entropy values of Chinese and American customers regarding popular aspects. From Table 12, it can be seen that Chinese customers' concerns about popular aspects are less homogeneous than those of American customers in all domains. Hence, we can conclude that concerns about popular aspects obviously differ between Chinese and American customers in all three domains, and the differences are much larger than those relating to frequent aspects. From Table 8(b), we can see that for digital camera reviews, the most popular aspect among Chinese customers is *genuine product*, followed by *logistics* and *price*, while American customers' most popular aspects are *quality, flashlight* and *battery*. Chinese customers are most unsatisfied with the *service* aspect, while American customers are most dissatisfied with *shutter*. From Table 9(b), we can see that for smartphone reviews, Chinese customers' most popular aspect is *genuine product*, followed by *price* and *logistics*, while American customers' most popular aspects are *quality, appearance* and *camera*. Both Chinese and American customers are dissatisfied with *apps*, while Chinese customers are most unsatisfied with *battery* and American customers are most dissatisfied with *apps*. From Table 10(b), we can see that for tablet computer reviews, Chinese customers' most popular aspect is *service*, followed by *genuine product* and *battery*, while American customers' most popular aspects are *quality, appearance* and *service*. Both Chinese and American customers are dissatisfied with *screen*, while Chinese customers are most dissatisfied with *screen* and American customers are most dissatisfied with *processor*.

**Table 12** Entropy values about popular aspects.

|          | Digital cameras | Smart phones | Tablet computers |
|----------|-----------------|--------------|------------------|
| Chinese  | 0.5549          | 0.7788       | 0.8361           |
| English  | 0.4269          | 0.6045       | 0.2887           |

The responses to the questionnaire survey show that the aspects of greatest concern to Chinese and American customers' differ, thus the aspects with which they are most satisfied or most dissatisfied are different. In addition, we can see that aspects that are of concern to customers in the two regions constitute a large proportion of the most frequent aspects, although their attitudes are not the same. For example, American customers are quite satisfied with *quality*, while Chinese customers are not. This suggests that manufacturers should strive to improve the quality of products in the Chinese market, especially in the tablet computer domain. For aspects with low satisfaction ratings in reviews from the two regions, enterprises need to pay more attention to customers' opinions in relation to their products. It is worth noting that Chinese customers show a high level of satisfaction with *price*, except in the tablet computer domain. This suggests that in the process of product promotion and marketing in China, price is not an important point, hence inputs for promoting product price can be reduced appropriately. However, in the tablet computer domain, Chinese customers are not satisfied with *price*. This may be because the popularity of the tablet computer is not sufficiently widespread. Smartphones have almost become a necessity of life, while digital cameras are required by photography enthusiasts. Quality information about smartphones has become quite comprehensive, therefore the price sensitivity is low [2], and most consumption of digital cameras is conspicuous consumption [5], which occurs when customers buy unnecessary or luxury products to display their financial resources and social status. Hence, consumers are willing to pay high prices for digital cameras. These two products are almost at two consumption extremes. The degree of quality information and level of conspicuous consumption for tablet computers lie somewhere in between these extremes, resulting in a low level of satisfaction regarding price.

(3) Extracting frequent aspect that are concerned both by Chinese and American customers

Frequent aspects that are subject to concern by both Chinese and American customers are shown in the following figures. Figure 5 shows frequent aspects of digital camera reviews, and it can be seen that there are six aspects that both Chinese and American customers are concerned about. *Screen* is the aspect with the lowest satisfaction level in both Chinese and English reviews. Chinese customers are not satisfied with *battery*, although American customers show high satisfaction levels. Attitudes toward *price* are reversed, with Chinese customers being more satisfied than American customers.

Figure 6 shows frequent aspects of smartphone reviews. As can be seen, there are seven aspects that both Chinese and American customers are concerned about. Satisfaction with *apps* is low in both Chinese and English reviews. Chinese customers are not satisfied with *battery*, while American customers show a high level of satisfaction. Again, their attitudes toward *price* are reversed, with Chinese customers being more satisfied than American customers.

Figure 7 shows frequent aspects of tablet computer reviews. As can be seen, there are eight aspects that both Chinese and American customers are concerned about. *Screen* is the aspect with the lowest satisfaction levels in both Chinese and English reviews. Chinese customers are not satisfied with *quality*, while American customers show a high level of satisfaction. Their attitudes toward *service* are reversed, with Chinese customers being more satisfied than American customers.

## 5.2 Customer behavior comparison results

By analyzing the responses to the questionnaire survey, we are able to compare customer behavior in the two regions. Corresponding with the questionnaire structure, there are three areas of customer behavior comparison: overall sentiment, brand preference, and purchase decision factors.

(1) Customer behavior comparison on sentiment expression

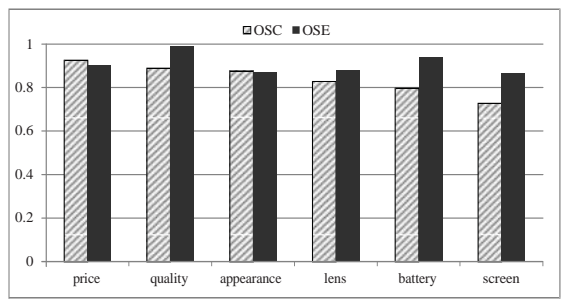

**Fig. 5** PA values of frequent aspects about digital cameras.

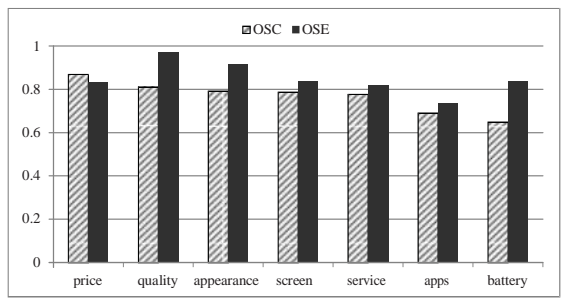

**Fig. 6** PA values of frequent aspects about smartphones.

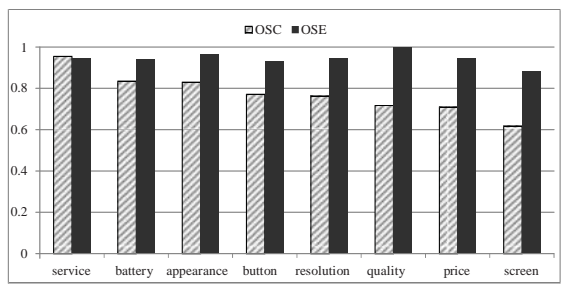

**Fig. 7** PA values of frequent aspects about tablet computers.

The results presented in Section 5.1.1, show that in relation to both overall satisfaction and product-level satisfaction, the OS values of Chinese customers are higher than those of American customers in relation to most conditions, hence we demonstrated that Chinese customers are more likely to express positive feelings. From the sociological point of view, Chinese people may be influenced by traditional culture, i.e., the *Doctrine of the Mean* [22]. This is an important part of Confucian philosophy, and involves an attitude of compromise, rather than leaning toward either side. The Doctrine of the Mean has been internalized by Chinese people for thousands of years, and is considered part of the national character. Hence, Chinese reviews include a wide variety of euphemisms, and they are accustomed to providing positive evaluations [58]. In contrast, Western culture encourages a more direct form of expression by American customers.

(2)  Customer behavior comparison on brand preference

The results presented in Section 5.1.2 show that American customers have more obvious BPs than Chinese customers. For example, because Samsung and Apple are both high-end smartphone brands, Chinese customers are not particularly concerned about specific aspects, and have no obvious BP; however, American customers prefer some aspects of Samsung smartphones. This may be related to differences in cultures or customers' experiences in different countries or regions [32, 50]. Based on our results, we can deduce that American customers pay more attention to details than Chinese customers in relation to IT products.

(3)  Customer behavior comparison on purchase decision factors

The results presented in Section 5.1.3 show that Chinese customers are concerned about the external features of products, such as the *logistics, package,* while American customers are concerned about product performance, including aspects, such as the *lens* and *quality*. From the sociological point of view, as a group, Chinese customers are more concerned about the external features of a product, while American customers pay more attention to the internal features. It is worth noting that *genuine product* is a frequently mentioned aspect in Chinese reviews. This suggests that the Chinese market should be subject to tight supervision regarding product quality to enhance customers' purchasing confidence. This supervision should come from both government and the manufacturing enterprises themselves. Additionally, multinational marketing needs to be guided in the light of its general trend, and different market conditions and customer behaviors should be taken into consideration.

## 5.3 A case for smartphone promotion

We now present an imaginary case using online reviews from China and America. Suppose that Peter is a staff member of a smartphone marketing company and wants to promote smartphones in China and America. As customers from different regions may have different preferences and concerns about different aspects regarding smartphones, it is necessary for product marketing staff to understand customer behavior in different regions. As a senior marketing staff member, Peter has received a brief report about the differences between Chinese and American customers, such as that shown in Table 13.

**Table 13** Example report of differences between Chinese and American customers.

| (a) Overall comparison | | |
|---|---|---|
| Comparisons | Chinese customers | American customers |
| Sentiment expression | euphemistic | direct |
| Brand preference | not obvious | obvious |

| (b) Aspect comparison |
|---|
| Purchase decision influencing factors |

Based on this report, Peter should create different marketing programs for customers in different regions.

First, regarding the matter of sentiment expression, Peter has taken the eu- phemistic expressions of Chinese customers into consideration. Therefore, he has raised the threshold for Chinese satisfaction that he obtained from test marketing. Second, Pe- ter is promoting an unpopular smartphone brand. When he promotes this smartphone to American customers, he focuses on contrasting this brand with popular brands, and emphasizing the advantages of this brand. Third, Chinese and American customers have different views in terms of the aspects that are of most concern, so Peter empha- sizes different aspects to different customers. When Peter promotes the smartphone to Chinese customers, he allocates most resources, including people, finance, and time, toward promoting the smartphone's external features. He chooses good logistics com- panies to ensure reliable, speedy delivery, and good e-commerce websites to ensure that the most important qualities of the smartphone are emphasized. When Peter pro- motes the smartphone to American customers, most of the resources are allocated to ensuring that the smartphone is of high quality. He emphasizes the need for software compatibility, high-definition screens, and long-lasting batteries.

Marketing based on the above analyses may improve customer satisfaction, which can in turn improve sales volume. Moreover, because of the targeted marketing pro- grams, marketing costs are likely to be greatly reduced.

## 6 Conclusion and future work

In this paper, we present a framework for developing an automatic questionnaire related to online customer behavior to explore differences between the online shopping behavior of Chinese and American customers. We generate questions and extract answers automatically via multi-granularity opinion mining. Three important conclusions can be drawn from our analysis:

(1) Chinese customers obey the Doctrine of the Mean and often use euphemistic expressions, while American customers use more direct expressions
(2) American customers pay more attention to product details than Chinese customers
(3) Chinese customers are more concerned about external features, while American customers are more concerned about internal features.

We investigate differences between Chinese and American online shopping behavior using reviews posted on e-commerce websites. This method can be extended to investigate differences between Chinese and American consumers on other websites, and differences between users from other countries or regions. In a word, this is a study about exploring differences in customers behavior by mining online reviews, which provide some reference values for research on features including user cognition, sentiment analysis.

The following future work is planned. First, this was our first attempt to generate questionnaires automatically, and only simple questions based on templates were generated. Therefore, we will conduct further research into areas such as question and answer technologies. Second, reviews from greater numbers of websites and regions will be compared and analyzed to determine whether differences in online shopping behaviors exist among customers who speak different languages and/or are from different continents. Third, we will conduct further research into sentiment analysis technologies to obtain more comprehensive sentiment results, and additional measurement techniques will be used to quantify differences in online shopping behaviors. Finally, these research topics can be expanded into other domains, such as catering, tourism, and entertainment.

## Acknowledgments

This work is supported by the Major Projects of National Social Science Fund (No. 13&ZD174), the National Social Science Fund Project (No. 14BTQ033), the Natural Science Foundation of China (No. 61305090, No. 61272233) and the Opening Foundation of Alibaba Research Center for Complex Sciences, Hangzhou Normal University (No. PD12001003002003). We thank the reviewers for the insightful comments.

## Compliance with Ethical Standards

The authors declare that they have no conflict of interest. All procedures performed in studies involving human participants were in accordance with the ethical standards of the institutional and/or national research committee and with the 1964 Helsinki declaration and its later amendments or comparable ethical standards. This article does not contain any studies with animals performed by any of the authors. Informed consent was obtained from all individual participants included in the study.